\documentclass[twocolumn]{article}
\usepackage{graphics}
\usepackage{latexsym}
\usepackage{epsfig}
\newtheorem{theorem}{Theorem}
\renewcommand{\deg}{^{\circ}}

\newcommand{\figtilemc}{4}

\begin{document}

\vspace*{-1.0in}
\begin{minipage}{\textwidth}
\noindent{\it\small This article appears in The Mathematical Intelligencer, Volume 29,
page 33 (2007).  The version printed there is slightly different.  Due
to a mix-up in the editorial process, it does not reflect a number of
changes made following the initial review.  Note especially the
addition of several references.}
\end{minipage}

\vspace{1.0in}
\hspace{0.6in}
\begin{minipage}{\textwidth}
{\bf\Huge \textsf{Hexagonal Parquet Tilings}} \\ 

\vspace{4pt}
{\bf\huge \textsf{{\it k}-Isohedral Monotiles}} \\ 

\vspace{-3pt}
{\bf\huge \textsf{with Arbitrarily Large {\it k}}}

\vspace{18pt}
{\bf\large \textsf{Joshua E.~S.~Socolar}} \\
\textsf{Physics Department and Center for Nonlinear and Complex Systems \\ 
 Duke University, Durham, NC 27708}

\end{minipage}

\vspace{0.6in}
The interplay between local constraints and global structure of
mathematical and physical systems is both subtle and important.  The
macroscopic physical properties of a system depend heavily on its
global symmetries, but these are often difficult to predict given only
information about local interactions between the components.  A rich
history of work on tilings of the Euclidean plane and higher
dimensional or non-Euclidean spaces has brought to light numerous
examples of finite sets of tiles with rules governing local
configurations that lead to surprising global structures.  Perhaps the
most famous now is the set of two tiles discovered by Penrose that can
be used to cover the plane with no overlap but only in a pattern whose
symmetries are incompatible with any crystallographic space group.
\cite{penrose,gardner} The Penrose tiles ``improved'' on previous
examples due to Berger \cite{berger} and others (reviewed by
Gr\"unbaum and Shephard \cite{GS}) showing that larger sets of square
tiles with colored edges (or several types of bumps and complementary
nicks) could force the construction of a non-periodic pattern.  

The discovery of a set of only two tiles that could fill space but
only in a non-periodic way raised a host of interesting questions.
The Penrose tilings have elegant geometric and algebraic
properties~\cite{penrose,debruijn,levine}.  One successful line of
research has been the discovery of 

\rule[0pt]{\linewidth}{0pt}

\vspace*{2.92in}
\noindent 
tile sets that have the Penrose
properties but different point group symmetries in
two~\cite{ammann,socolar2,socolar3} and three
dimensions\cite{ammann,socolar1} or in hyperbolic space
~\cite{goodman-strauss}.  In all 
of these cases, the rules one must
follow to construct a tiling are strictly local.  Any configuration in
which adjacent tiles fit together to leave no holes is allowed.  There
is no explicit constraint on the relative positions of tiles that do
not touch each other.

Another question, which has proven more difficult, is the quest for a
single tile (rather than a set of two) that forces a non-periodic,
space-filling tiling of the plane.  It may be fruitful to view this as
a limiting case of the following more general problem.  Any tiling can
be classified according to its isohedral number $k$, defined as the
size of the largest set of tiles for which no two can be brought into
coincidence by a global symmetry (any reflection, rotation,
translation, or any combination of these) that leaves the entire
tiling invariant.  A set of tiles for which the smallest isohedral
number of an allowed tiling is $k$ is called a {\it k-isohedral set}.
If the set consists of a single tile, the tile is called a
$k$-isohedral ``monotile.''  The challenge is to find a $k$-isohedral
monotile for arbitrarily large $k$.

\begin{figure*}[t]
\centering
\includegraphics[clip,width=0.6\linewidth]{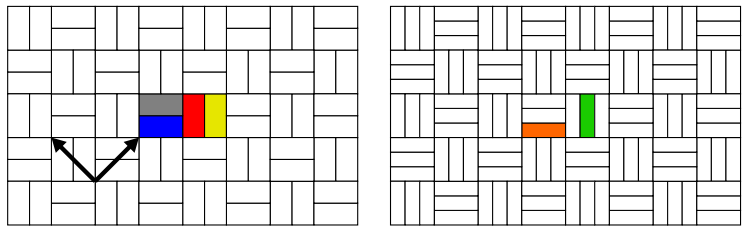}
\caption{A 1-isohedral tilings and a 2-isohedral tiling.}
\label{fig:k12}
\end{figure*}

To gain some intuition about the isohedral number, consider the two
tilings shown in Fig.~\ref{fig:k12}.  The tiling on the left has
$k=1$; any tile can be mapped to any other by a translation that
leaves the entire tiling invariant.  The red tile can be mapped into
the yellow one by a $180\deg$ rotation about the midpoint of their
common edge; into the blue one by a counterclockwise $90\deg$ rotation
about the lower left corner of the red tile; and into the gray one by
a clockwise $90\deg$ rotation about the upper left corner of the red
tile.  Combining these rotations with the square lattice of
translations generated by the vectors shown allows any tile to be
mapped into any other.  The tiling on the right has $k=2$.  It has the
same symmetries as the one on the left, but there is no symmetry that
maps the green tile into the orange one; there are two ``inequivalent
types'' of tiles in this tiling.

The answer to the question ``Is there a $k$-isohedral monotile?''  for
arbitrarily large $k$ depends crucially on how the question is posed.
As we will see below, there are many subtly different versions of this
and similar questions, and versions that may at first glance appear
equivalent turn out not to be.  Forcing nontrivial global structure of
a certain precisely defined type can be accomplished in a variety of
ways depending on what types of local {\it matching rules} are deemed
permissible.  We will assume that the rules are applicable only to
tiles that share (some portion of) an edge.  But shall we require that
the monotile be completely defined by its shape alone? Or shall we
allow coloring of the edges and specification of which colors are
allowed to coincide?  Shall we insist that the monotile be a simply
connected shape?  Shall we insist that the tiling cover the entire
plane, or just that it have the highest possible density?

Recent exhaustive searches of polyomino monotiles consisting of
square, triangular, or hexagonal units have produced $k$-isohedral
examples with $k$ as large as 10 (so far!) \cite{myers}, but there
appears to be no systematic way construct such examples analytically.
In these examples, the matching rules are enforced by shape alone and
the entire space must be covered.

\begin{figure}[t]
  \centering
  \includegraphics[clip,width=0.7\linewidth]{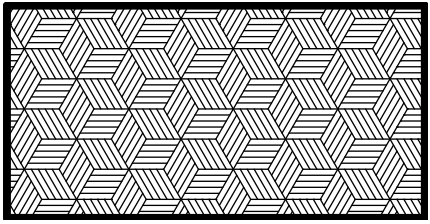}
\caption{The $L=5$ hexagonal parquet tiling of the plane.  Each ``board'' is 
  a copy of the same tile.}
\label{fig:hexparquet}
\end{figure}

Below we present several variations of a class of monotiles and
matching rules that can force tilings with arbitrarily large $k$.  The
tilings formed all have the basic structure of the {\em hexagonal
  parquet} shown in Fig.~\ref{fig:hexparquet}.  Each rhombus in the
figure is composed of $L=5$ monotiles.  Generalization to arbitrarily
large $L$ is clearly possible.
(Exercise: Find the isohedral number of the hexagonal parquet in terms
of $L$.  Answer below.)

In each of the following four sections, we present a monotile and
matching rule that forces a tiling with the symmetry of the hexagonal
parquet.  The difference between the tilings lies in the way in which
the rule is expressed.  Defining a tile to be the closed set of points
bounded by the tile edges (and faces in higher dimensions), we have
the following four cases.  In all cases, we allow tiles to overlap
only along edges.
\begin{enumerate}
\item The edges of the monotile are colored, there are rules
  constraining which colors can coincide, and the tiles cover the
  entire space.
\item The edges are not colored.  The rule is that the tiles must
  cover the entire space, but the tile is not a simply connected
  shape.
\item The monotile is simply connected and the rule is that the tiling
  must maximize the density of tiles without necessarily covering the
  entire space.
\item The (3D) monotile is a simply connected shape and the only rule
  is that the tiles must fill the space (or just an infinite slab
  thick enough to accommodate the height of one tile).
\end{enumerate}
In this work we allow only rotations and translations of the monotile,
not reflections.  All of the results can be easily extended to the
case where reflections are allowed by replacing the disks, bumps, and
nicks with chiral shapes.

Perhaps just as important as the discovery of monotiles that force any
desired isohedral number, these examples show that subtle differences
in the rules of the game may generate dramatically different results.
Note that we have {\it not} exhibited a simply connected, uncolored,
two-dimensional monotile that forces the hexagonal parquet structure.
In fact, we show below that this is impossible.  If you want to
restrict the problem to these terms, the record in two dimensions is
still Myers' polyomino consisting of a simply connected cluster of 16
hexagons.~\cite{myers}

\vspace{8pt}
\noindent{\bf\large\textsf{A 2D monotile with color matching rules}}

\vspace{2pt}

\begin{figure}[t]
\centering
\includegraphics[clip,width=0.73\linewidth]{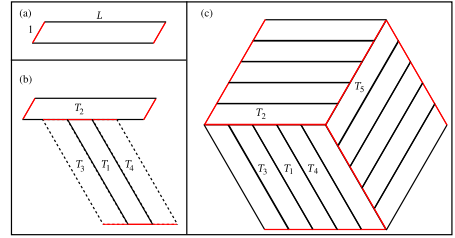}
\caption{The $L=5$ hexagonal parquet monotile: (a) the tile; (b) forced tiles; (c) the forced hexagon.}
\label{fig:tile1}
\end{figure}

The tile shown in Fig~\ref{fig:tile1}a is endowed with a matching rule
requiring that no two red edges may touch.  As the aspect ratio of the
tile is increased in integer steps, the minimal isohedral number of a
space-filling tiling formed with this tile increases without bound.

\vspace{-0.2in}
\begin{theorem}
  Let $T$ be a parallelogram tile with angles of $60^{\deg}$ and
  $120^{\deg}$ and side lengths $1$ and $L>1$, where $L$ is an
  integer.  (See Fig.~\ref{fig:tile1}a.)  Color the short edges of $T$
  red (not including the vertices) and the long edges (and vertices)
  black.  The minimal isohedral number of a tiling in which no points
  are covered twice with red is $\lfloor (L+1)/2 \rfloor$.
\end{theorem}

\noindent{\bf Proof:} Consider the tile $T_1$ shown in Fig.~\ref{fig:tile1}b.
The matching rule and requirement of space filling immediately imply
that $T_2$ must be present.  The only way to continue the tiling is
then to place $T_3$ and $T_4$ as shown with dashed outlines.  The
process of adding forced tiles stops only when the ends of $T_2$ are
reached through further additions of tiles along its bottom edge.  At
this point, $T_5$ in Fig.~\ref{fig:tile1}c is forced and the process
repeats until the hexagon of Fig.~\ref{fig:tile1}c is formed.

The existence of this hexagon in the tiling ensures that the isohedral
number of the tiling is at least $\lfloor (L+1)/2 \rfloor$.  Each tile
can be characterized by the distance of its center from the center of
the rhombus containing it, with tiles on opposite sides of the center
possibly related by rotation of $180^{\deg}$ about the center.

By inspection it is clear that the hexagons can tile the plane while
respecting the matching rules, forming a standard honeycomb lattice in
which the isohedral number is exactly $\lfloor (L+1)/2 \rfloor$.  
\hfill $\Box$

\begin{theorem}
  For $L>2$, the color matching rule for the monotile $T$ cannot be
  enforced by alterations of the tile shape alone; i.e., by placing
  bumps and nicks on the tile edges.
\end{theorem}

\noindent{\bf Proof:} 
Let the shape of one red edge be designated $R_1$ and the shape of the
other red edge be $R_2$.  Further, let $R_1'$ and $R_2'$ be the
complementary shapes that fit onto $R_1$ and $R_2$, respectively.  The
color matching rule implies that neither $R_1$ nor $R_2$ is congruent
to $R_1'$ or $R_2'$.  We will now show that the number of instances of
$R_x$ on the tile must be greater than the number of instances of
$R_x'$, which immediately implies that $T$ cannot tile the plane.

\noindent
\begin{minipage}{\textwidth}
\begin{center}
\includegraphics[clip,width=0.4\linewidth]{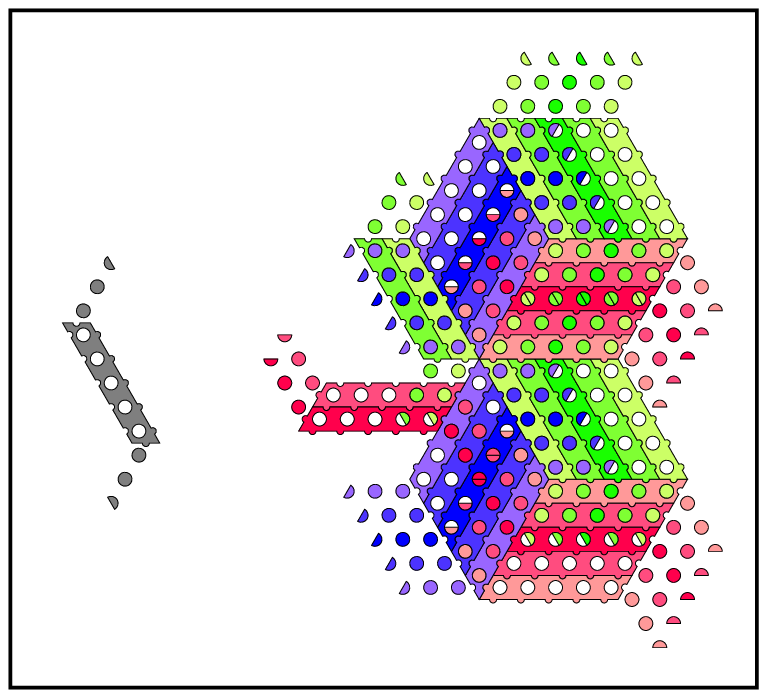} \hspace{0.\linewidth}
\includegraphics[clip,width=0.4\linewidth]{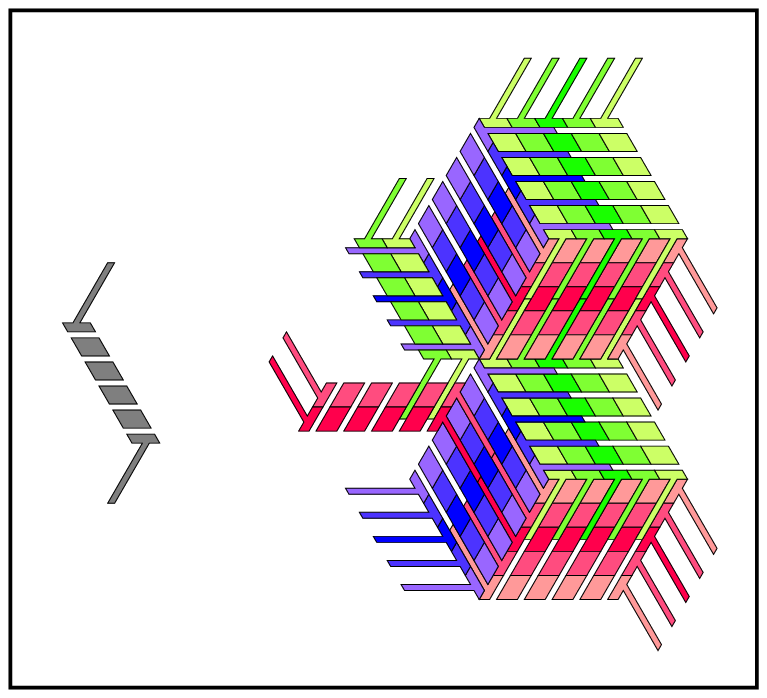}
\end{center}
  Figure 4:
  An $L=5$ multiply connected monotile (gray) that forces the
  hexagonal parquet tiling.  Colors are guides to the eye to help
  identify individual tiles.
\addtocounter{figure}{1}
\end{minipage}

\vspace{6pt}
Consider any tile in the interior of a rhombus, which has both black
edges matching black edges of its neighbors.  Let the shapes of the
two black edges be $B_1$ and $B_2$.  There are two possibilities: (1)
$B_1 = B_2'$, which implies $B_2 = B_1'$; or (2) $rB_1 = B_1'$ and
$rB_2 = B_2'$, where $rX$ indicates rotation of $X$ by $\pi$.  In
either case, any instance of $R_x'$ found on $B_1$ must be matched by
an instance of $R_x$ either on $B_2$ or on $rB_1$.  Thus the number of
instances of $R_x'$ on black edges cannot exceed the number of
instances of $R_x$.  This means that the single $R_x$ on the red edge
makes the number of $R_x$'s larger.  The conclusion is that in order to
tile the plane, the red edge matching rule has to be relaxed, but this
in turn permits simple periodic tilings with isohedral numbers or one
or two. \hfill $\Box$

\vspace{8pt}
\noindent{\bf\large\textsf{Forcing the hexagonal parquet with a multiply connected monotile}}

\vspace{2pt}

The color matching rule for the hexagonal parquet monotile can be
enforced by shape alone if one does not insist on $T$ being simply
connected.  The proof is by construction, as displayed in
Fig.~\figtilemc.  The seven black regions at the left of the
figure form the monotile.  By inspection, it is clear that there is no
way to have two short edges of the basic parallelogram coincide.  (The
nearby disks or protruding rods get in the way.)  Thus the rules for
how the parallel-

\rule[0pt]{\linewidth}{0pt}

\vspace*{2.74in}
\noindent
ograms can be placed are at least as restrictive as
the color matching rules discussed above.  The figure clearly shows,
however, that the hexagonal parquet tiling can still be formed.

\vspace{8pt}
\noindent{\bf\large\textsf{Forcing the hexagonal parquet with a simply connected 3D monotile}}

\vspace{2pt}

\begin{figure}[b]
\centering
\includegraphics[clip,width=0.3\linewidth]{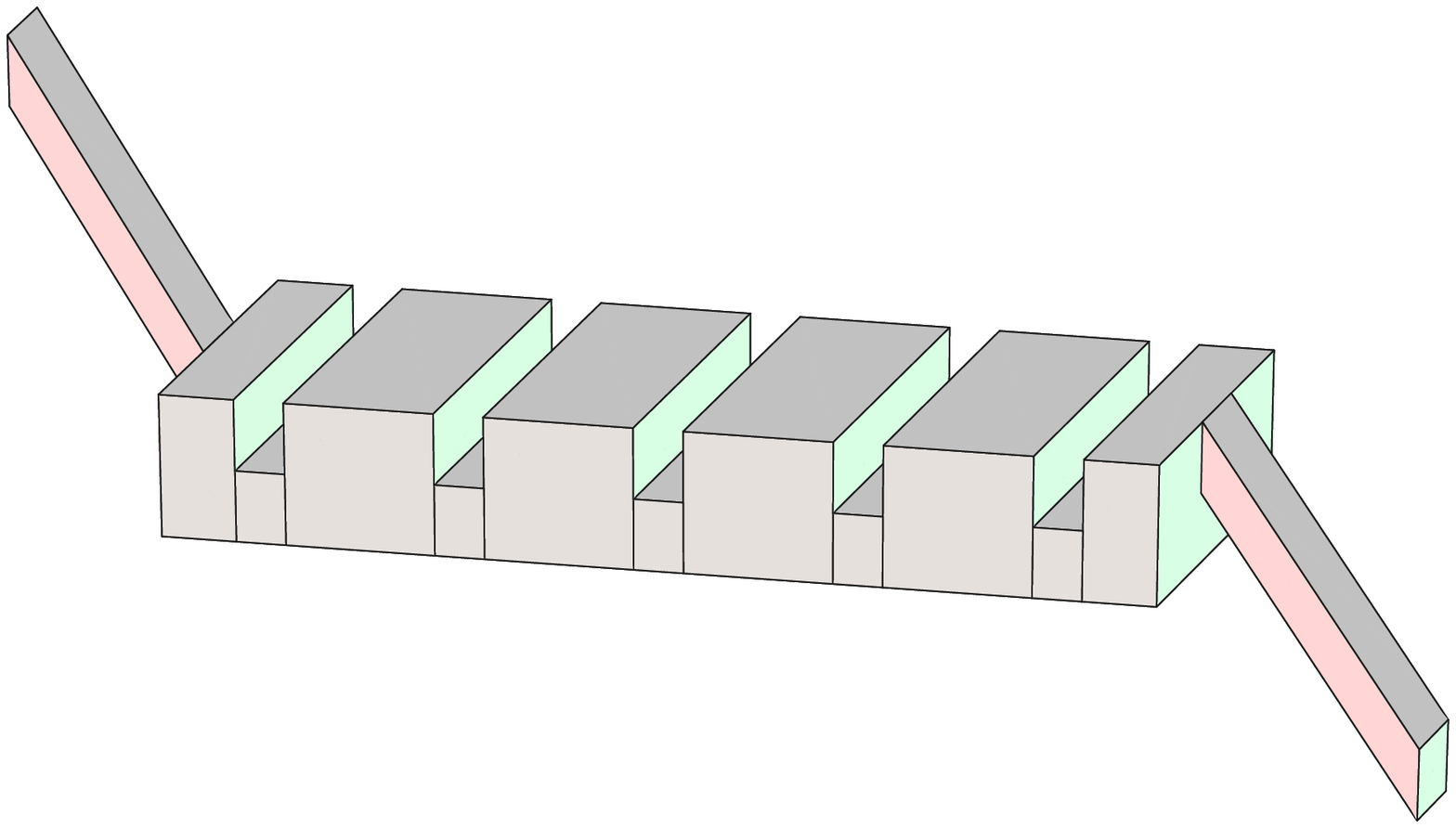}
\includegraphics[clip,width=0.3\linewidth]{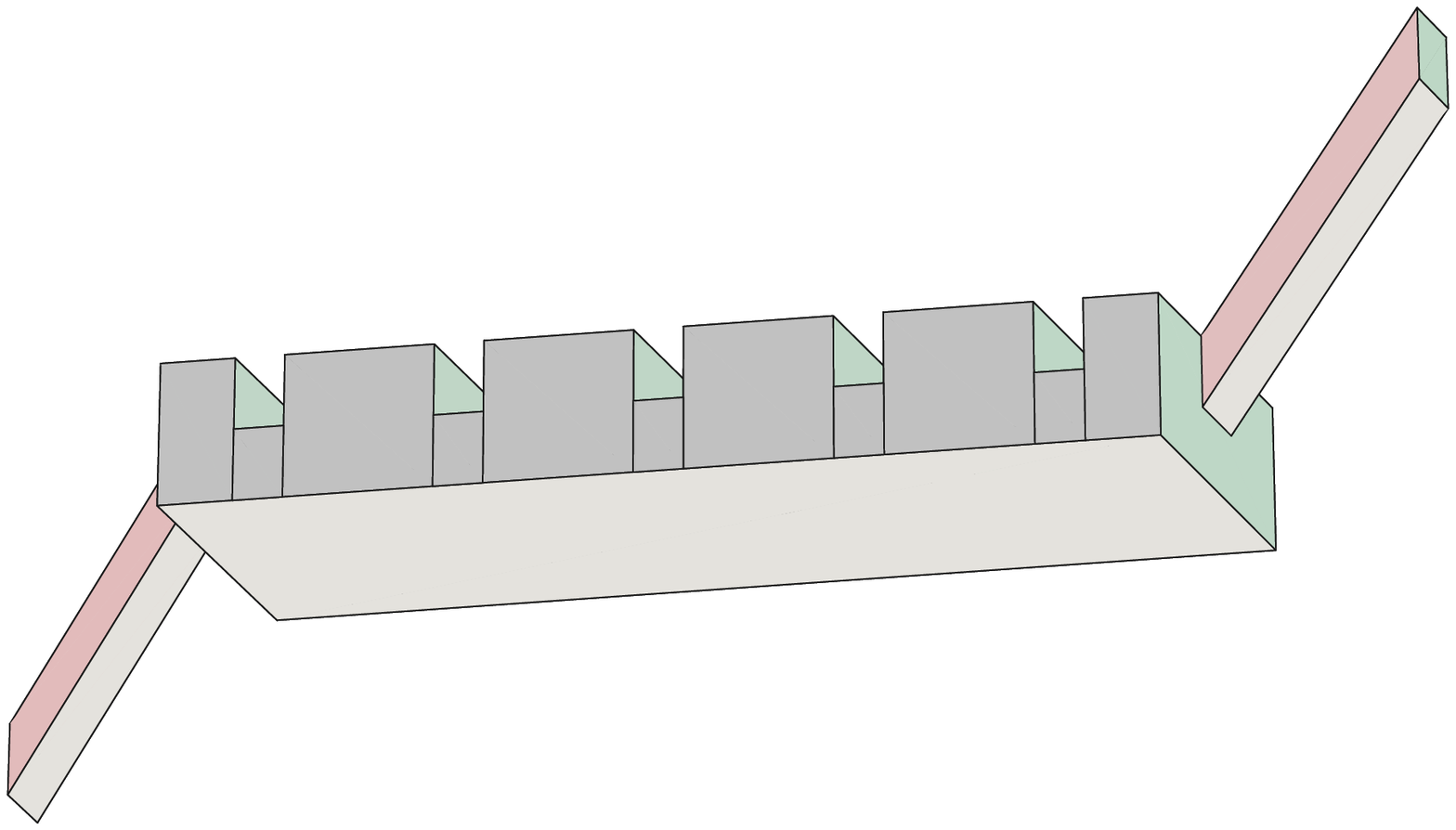}
\includegraphics[clip,width=0.3\linewidth]{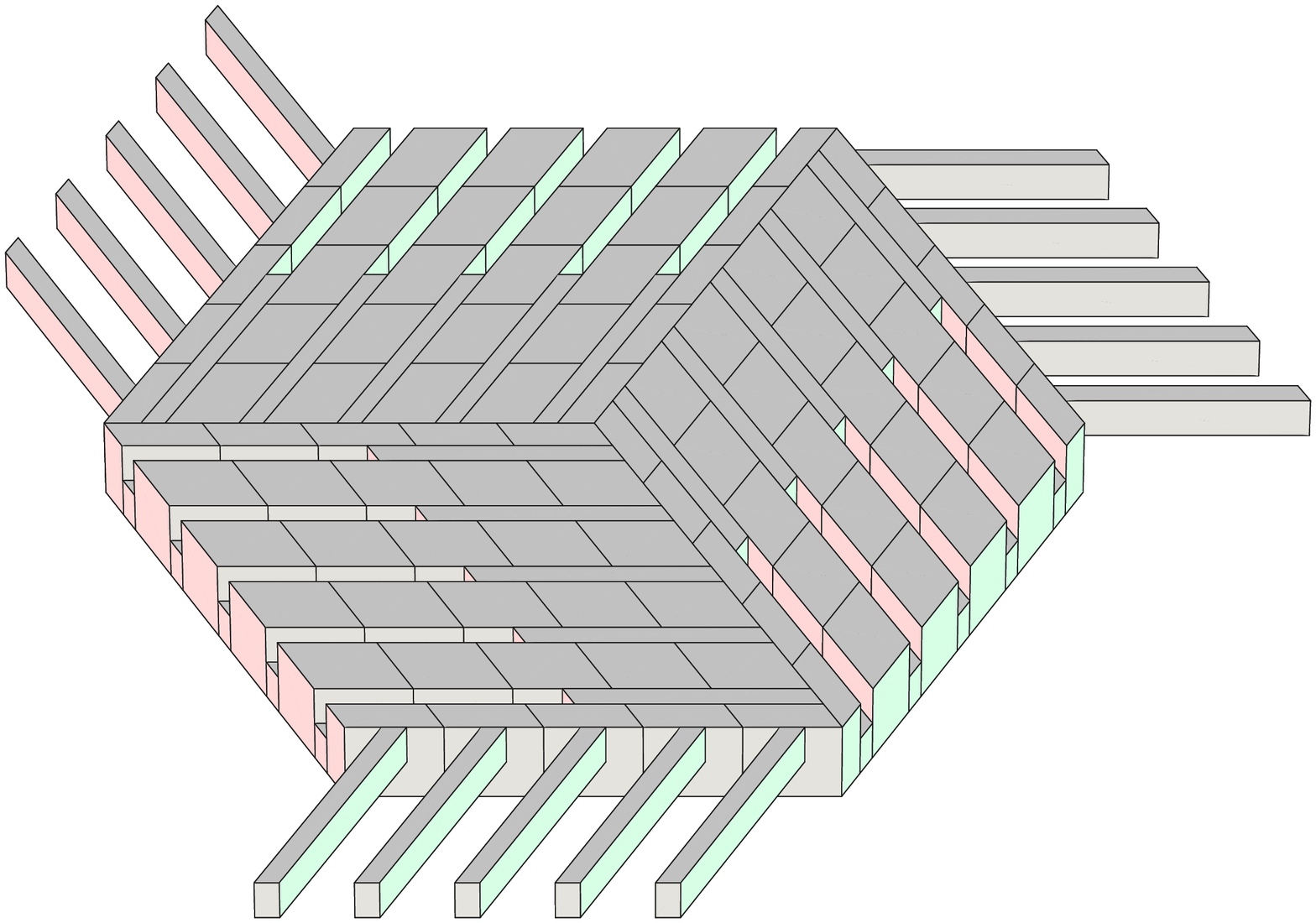}
\caption{
  A simply connected monotile that forces hexagonal parquet layers
  that can be stacked to fill space.  Top, bottom, and tiling views.}
\label{fig:tile3D1}
\end{figure}

The color matching rule required for the hexagonal parquet tile can
also be implemented with a simply connected monotile in three
dimensions.  The simplest way to do it is to promote the multiply
connected 2D monotile on the right in Fig.~\figtilemc\ to a 3D
parallelepiped with shallow protruding rods and grooves as shown in
Fig.~\ref{fig:tile3D1}.  The complete tiling is a stacking of
identical hexagonal parquet layers.  The lowest permitted isohedral
number for the space-filling 3D tiling is the one in which the layers
are in perfect registry.  This can be forced, if desired, by placing
bumps on the rods at the positions corresponding to the disk centers
in the monotile of the left panel of Fig.~\figtilemc\ and
corresponding dents in the bottom of the parallelepiped.  Note that
the pattern of disks in Fig.~\figtilemc\ is {\em not} a
triangular lattice, so the registry is indeed forced.

The multiply connected tiling on the left in Fig.~\figtilemc\
suggests a different strategy for constructing a 3D monotile.  The
tiling now consists of stacks of double layers, each double layer
being a hexagonal parquet with flat top and bottom surfaces.  The
enforcement of the matching rule for the top of the double layer is
provided by the pieces of tile on the bottom of the double layer.  The
protrusions and indentations on the top surface of the bottom-layer
pieces do not fit properly into those in the top-layer piece when one
attempts to match the top pieces end to end.  Thus one is forced to
form a hexagonal parquet in a manner quite similar to the multiply
connected 2D tiling above, with the bottom-layer pieces playing
exactly the same role as the isolated disks in the 2D monotile.

\begin{figure}[t]
\centering
\includegraphics[clip,width=0.48\linewidth]{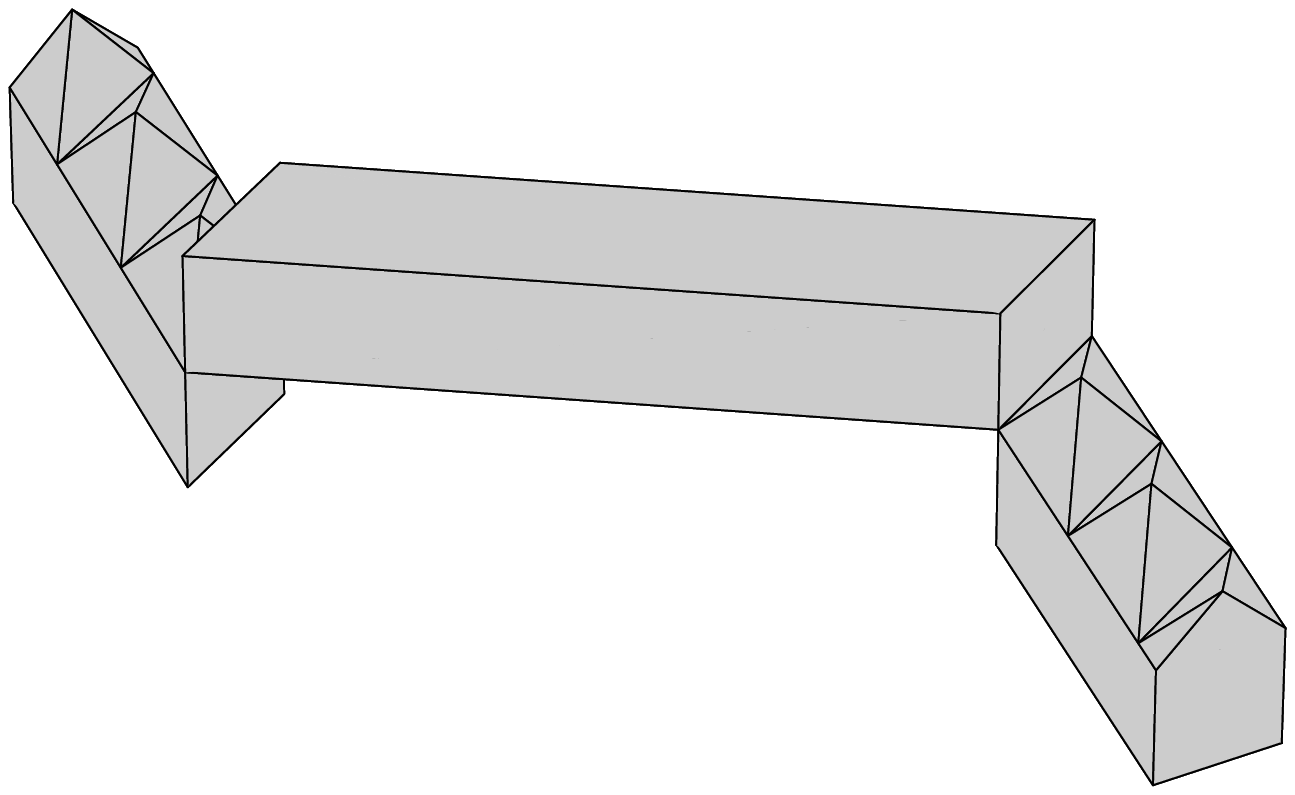}
\includegraphics[clip,width=0.48\linewidth]{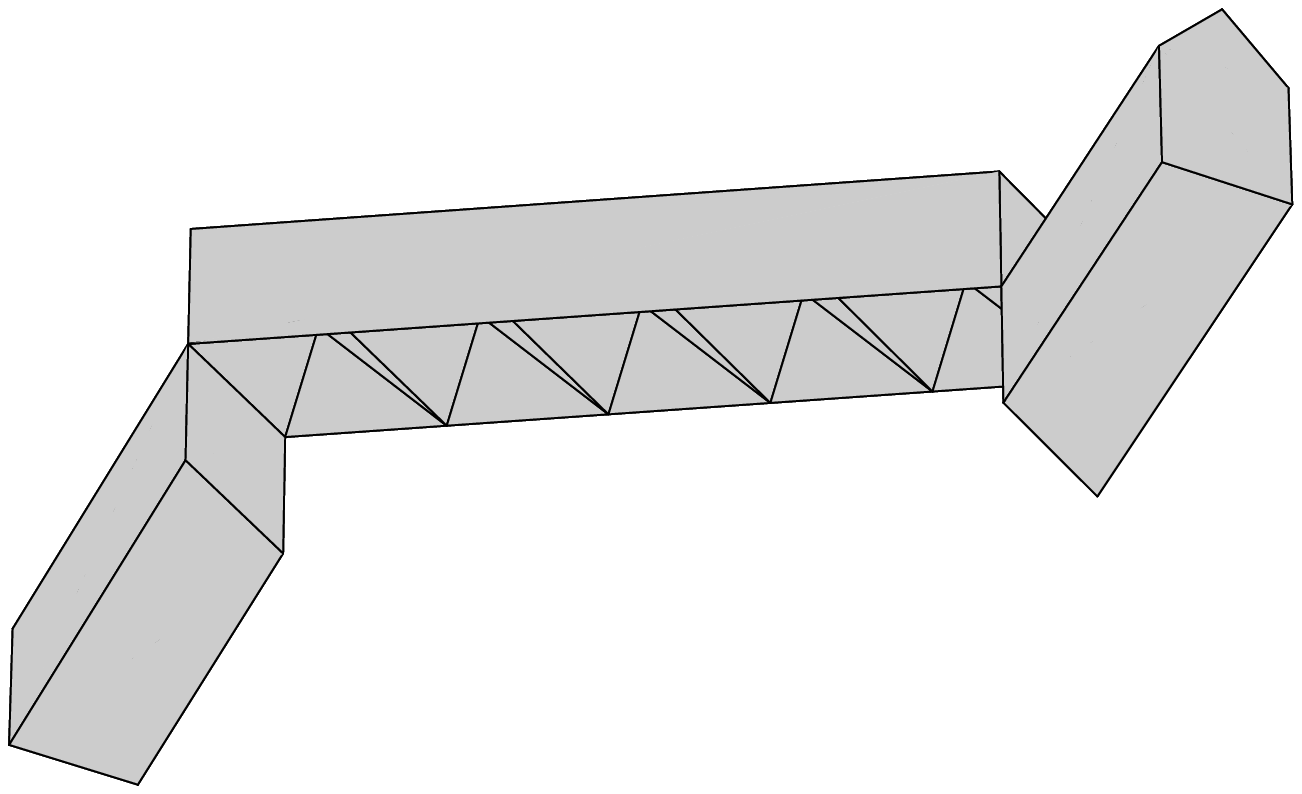}
\includegraphics[clip,width=0.48\linewidth]{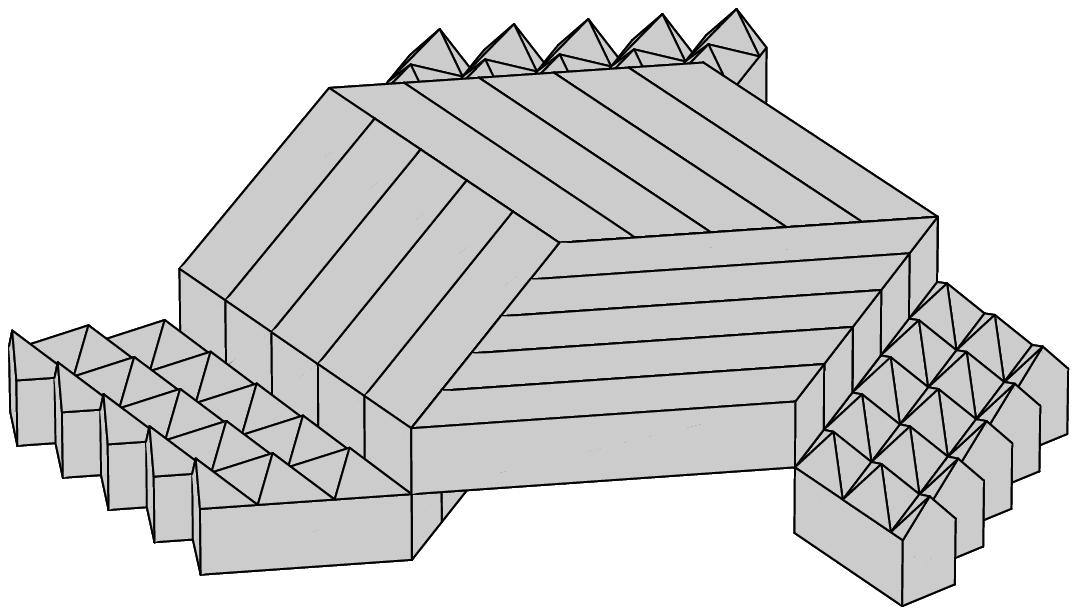}
\includegraphics[clip,width=0.48\linewidth]{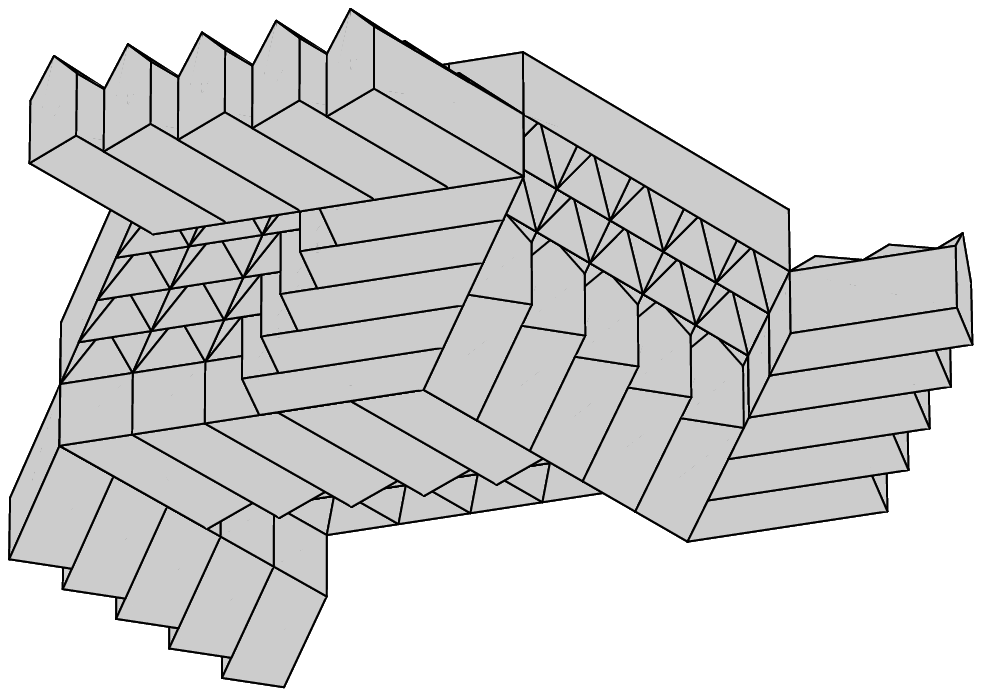}
\caption{
  A monotile that forces a double-layered hexagonal parquet and a unit
  cell of space-filling tiling.  Top and bottom views.}
\label{fig:3D2}
\end{figure}

One realization of this 3D monotile and one unit cell of the double
layer are shown in Fig,~\ref{fig:3D2}, each being shown from viewpoints
above and below the plane of the double layer.

\begin{figure}[t]
\centering
\includegraphics[clip,width=0.48\linewidth]{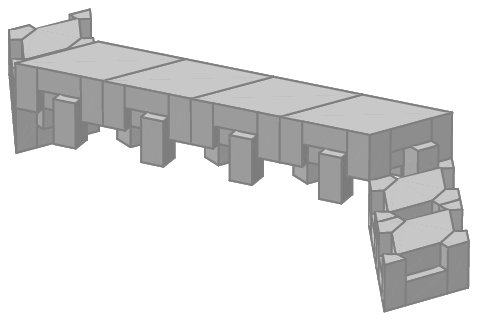}
\includegraphics[clip,width=0.48\linewidth]{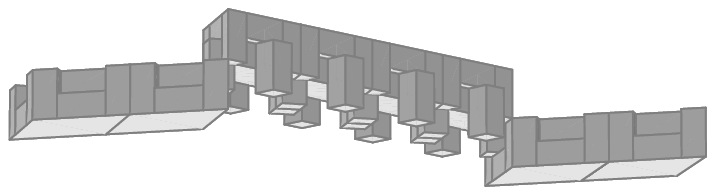}
\caption{
  A simply connected monotile that forces a double-layered hexagonal
  parquet.} 
\label{fig:3Dsc}
\end{figure}

The monotile of Fig.~\ref{fig:3D2} would not be simply connected if we
took the tile to be the open set not containing edges.  The
construction can be modified, however, so as to make even this open
set simply connected.  The $L=4$ version of the modified tile is shown
in Fig.~\ref{fig:3Dsc}.  The protruding ``legs'' from the top-layer
piece will fit into the grooves in the bottom-layer piece, with two
legs (one from each of two neighboring tiles) filling each hole formed
by neighboring tiles on the bottom layer.  The legs protruding from
the end of the top-layer piece and fitting into half of the groove on
the end of each bottom-layer piece form a connection that makes the
whole tile simply connected.

\vspace{8pt}
\noindent{\bf\large\textsf{Forcing the hexagonal parquet with a maximum density rule}}

\vspace{2pt}

\begin{figure}[t]
\centering
\includegraphics[clip,width=0.6\linewidth]{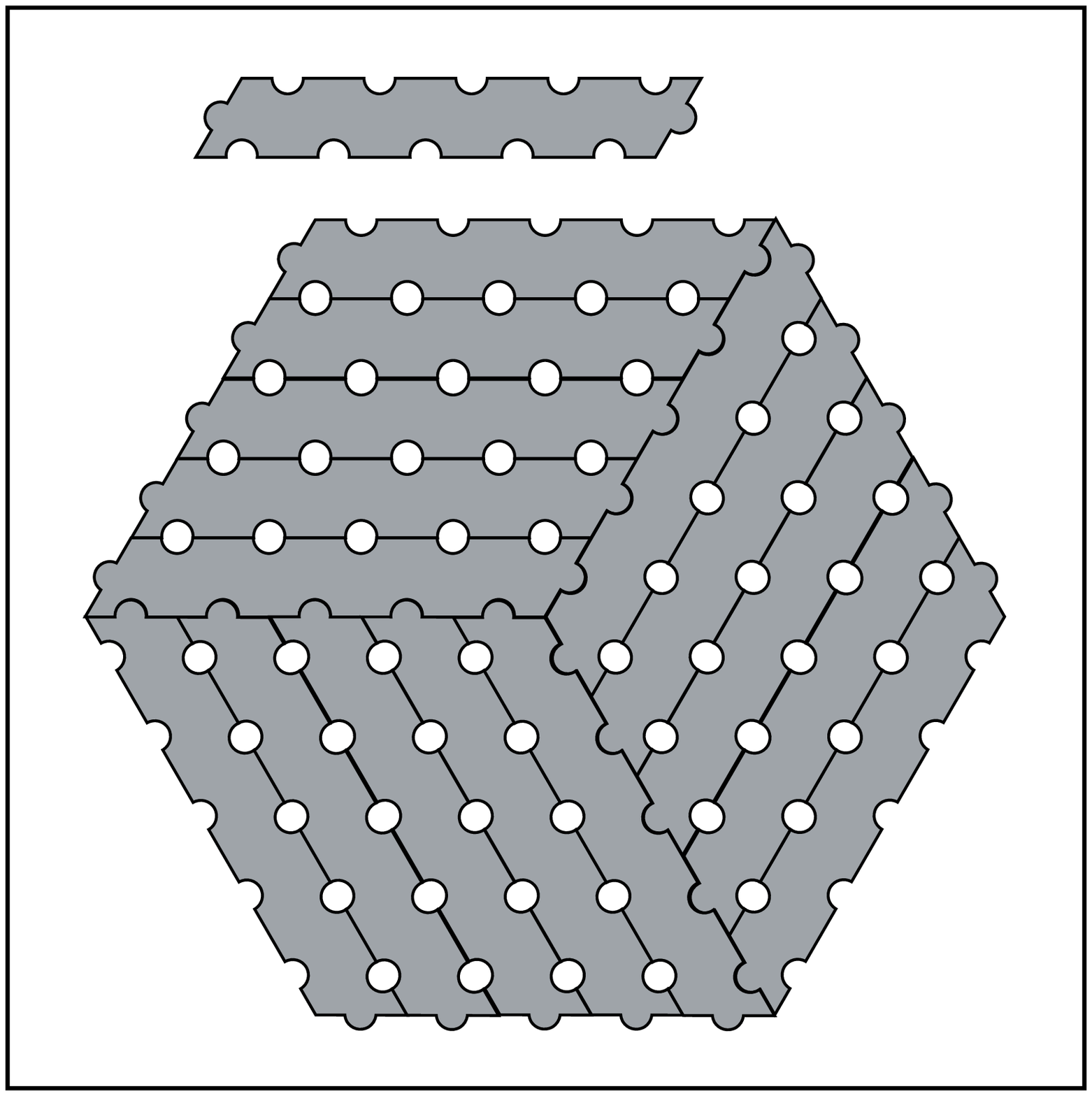}
\caption{
  An $L=5$ monotile (top) for which the hexagonal parquet is the
  maximum density tiling.}
\label{fig:tile3}
\end{figure}

The hexagonal parquet can be enforced by a simply connected shape in
2D if one replaces the space-filling constraint with the demand that
the tiling have the maximum possible tile density.  The shape in
Fig.~\ref{fig:tile3} can form a hexagonal parquet tiling as shown.
The color matching rule is enforced by the bumps on the ends of the
tile.  The parquet tiling is then the maximum density tiling that can
be achieved with this tile.  Because the smallest excluded area around
a tile occurs when its ends are fitted into notches, every tile in the
parquet tiling excludes the smallest area possible.

\vspace{8pt}
\noindent{\bf\large\textsf{Conclusions}}

\vspace{2pt}

We have exhibited several types of monotiles with matching rules that
force the construction of a hexagonal parquet.  The isohedral number
of the resulting tiling can be made as large as desired by increasing
the aspect ratio of the monotile.  Aside from illustrating some
elegant peculiarities of the hexagonal parquet tiling, the
constructions demonstrate three points:
\begin{enumerate}
\item Monotiles with arbitrarily large isohedral number do exist;
\item The additional topological possibilities afforded in 3D allow
  construction of a simply connected monotile with a rule enforced by
  shape only, which is impossible for the hexagonal parquet in 2D;
\item The precise statement of the tiling problem matters --- whether
  color matching rules are allowed; whether multiply connected shapes
  are allowed; whether space-filling is required as opposed to just
  maximum density.
\end{enumerate}

So what about the quest for the $k=\infty$ monotile?  Schmitt, Danzer
and Conway have exhibited a 3D monotile that forces a non-periodic
tiling.~\cite{danzer,baake,radin} The tiling is a stacking of
identical layers and each layer is a periodic packing of the monotile.
The non-periodicity arises because the planar lattice directions in
successive layers are rotated by an angle incommensurate with $2\pi$.
This tiling has an unusual feature: the number of local configurations
around a monotile is infinite.  That is, no two tiles in a given layer
are covered in exactly the same way by the tiles in the layers above
and below it.  In fact, the layers can slide over each other to form
an infinite number of tilings that are not related by any global
symmetry.  Any attempt to enforce a finite set of local environments
for this monotile will require a commensurate rotation angle and
render the isohedral number finite, though it could be arbitrarily
large.

Another example of a $k=\infty$ monotile is the decagonal tile
together with matching rules allowing certain types of overlap first
presented by Gummelt.~\cite{gummelt}  Jeong and Steinhardt proved that
the overlap rules and the requirement that the tile density be
maximized force a structure with the same symmetries as the Penrose
tiling.~\cite{jeong}  

At present there is there no general theory distinguishing patterns
that can be enforced by color matching rules from those that can be
enforced by shape alone or by maximum density constraints.  The
maximum density criterion is of particular interest in physics -- and
is particularly vexing because of the difficulty of linking this
global criterion to local constraints that can be exhaustively
checked.  In some cases, such as the hexagonal parquet case above, it
can be proven that satisfying certain local constraints will guarantee
maximum density.  The recent proof that the FCC packing of spheres in
three dimensions has maximum density is another example.~\cite{FCC} On
the other hand, there is some evidence that the maximum density sphere
packing in many dimensions is actually a random packing
\cite{torquato}, which would have an infinite isohedral number and an
infinite number of local configurations around a single sphere
(monotile).

As the examples described above suggest, there may be surprisingly
simple links between local rules and global structure, and the
collection of interesting specimens has yet to be placed within a
well-defined theoretical framework.  Tiling enthusiasts around the
world are looking for new ideas and examples that will lead to deeper
understanding, enjoying the recreational nature of the puzzles that
crop up, and appreciating the visual and logical structures that
emerge along the way.

\vspace{8pt}
\noindent{\bf\textsf{Acknowledgments}}

\vspace{2pt}

I thank C.~Goodman-Strauss and M.~Senechal for
their generous mathematical and editorial advice.

\onecolumn
\bibliographystyle{unsrt}

\end{document}